\documentclass[a4paper,twocolumn]{IEEEtran}
\usepackage{amssymb}
\usepackage{pifont}
\usepackage{booktabs}
\usepackage{authblk}
\usepackage{balance}
\usepackage{url}
\usepackage{booktabs,subcaption,amsfonts,dcolumn}
\usepackage{silence}
\usepackage{amssymb}
\renewcommand{\baselinestretch}{1}
\usepackage[font=scriptsize]{caption}
\let\labelindent\relax
\usepackage[pdftex]{graphicx}
\usepackage{epstopdf}
\usepackage{pifont}
\usepackage{graphicx}
\usepackage[dvipsnames]{xcolor}
\usepackage{array}
\usepackage{verse}
\usepackage{longtable}
\usepackage{comment}
\usepackage{supertabular}
\usepackage{cite}
\usepackage{amsmath}
\usepackage{lettrine}
\usepackage{hyperref}
\usepackage{multirow}
\usepackage{dirtytalk}
\usepackage[euler]{textgreek}
\newcommand{\RNum}[1]{\uppercase\expandafter{\romannumeral #1\relax}}
\usepackage{graphicx}
\usepackage[a4paper,margin=0.9in]{geometry}
\usepackage[ruled,vlined]{algorithm2e}
\usepackage{graphicx}
\usepackage{caption}
\usepackage{subcaption}
\usepackage{algorithmic}

\hypersetup{
    colorlinks=false,
    linkcolor=blue,
    filecolor=magenta,
    urlcolor=cyan,
}

\begin{document}

\title{Intelligent Resource Allocation in Dense LoRa Networks using Deep Reinforcement Learning}
\author{Inaam Ilahi$^1$}
\author{Muhammad Usama$^{1,2}$}
\author{Muhammad Omer Farooq$^3$}
\author{Muhammad Umar Janjua$^1$}
\author{Junaid Qadir$^{1,4}$}
\affil{$^1$ Information Technology University (ITU), Punjab, Lahore, Pakistan}
\affil{$^2$ Lahore University of Management Sciences (LUMS), Pakistan}
\affil{$^3$ Department of Systems and Computer Engineering, Carleton University, Canada}
\affil{$^4$ Department of Computer Science and Engineering, College of Engineering, Qatar University, Doha, Qatar}

\maketitle
\thispagestyle{empty}

\begin{abstract}
\label{abstract}
The anticipated increase in the count of IoT devices in the coming years motivates the development of efficient algorithms that can help in their effective management while keeping the power consumption low. In this paper, we propose \textcolor{black}{an intelligent multi-channel resource allocation algorithm for dense LoRa networks termed as }LoRaDRL and provide a detailed performance evaluation. Our results demonstrate that the proposed algorithm not only significantly improves LoRaWAN's packet delivery ratio (PDR) but is also able to support mobile end-devices (EDs) while ensuring lower power consumption \textcolor{black}{hence increasing both the lifetime and capacity of the network.} Most previous works focus on proposing different MAC protocols for improving the network capacity, i.e., LoRaWAN, delay before transmit etc. We show that through the use of LoRaDRL, we can achieve the same efficiency with ALOHA \textcolor{black}{compared to LoRaSim, and LoRa-MAB} while moving the complexity from EDs to the gateway thus making the EDs simpler and cheaper. Furthermore, we test the performance of LoRaDRL under large-scale frequency jamming attacks and show its adaptiveness to the changes in the environment. We show that LoRaDRL's output improves the performance of state-of-the-art techniques resulting in some cases an improvement of more than 500\% in terms of PDR compared to learning-based techniques.
\end{abstract}


\section{Introduction}
\label{sec:introduction}


The count of Internet-of-Thing (IoT) devices is anticipated to increase manifold in the coming years. These non-uniformly distributed dense networks will include end-devices (EDs) moving with different velocities. This puts forward a need for effective algorithms able to manage all those devices while keeping the collisions and the energy consumption as low as possible. Long-Range (LoRa) is a leading Low Power Wide Area Network (LPWAN) technology and \textcolor{black}{LoRaWAN is a leading LPWA networking protocol for LoRa}. LoRa uses the Chirp Spread Spectrum (CSS) technique which: (i) is resilient to interference, (ii) uses low power, (iii) is resistant to multi-path fading, (iv) is resistant to the Doppler effect, and (v) has low communication link budget \textcolor{black}{\cite{reynders2016chirp}}. LoRaWAN networks need a small infrastructure to be deployed and their scalability can be increased by adding more gateways to the network. This makes LoRaWAN an attractive low-cost IoT solution for transmitting data from the ED to the user and control commands from the user to the ED.

Broadly speaking, there are two critical factors that decide the usefulness of LPWAN: (i) better lifetime; and (ii) network capacity (i.e., the maximum number of EDs supported by the network). Battery lifetime is affected by the number of transmissions and the PHY-layer parameters used for transmission while the network capacity is affected by (i) the number of available channels, (ii) air time \textcolor{black}{(the time taken in air by the signal to reach the receiver)}, (iii) inter-transmission time, and (iv) transmission power. \textcolor{black}{Dynamic allocation of PHY-layer parameters in LoRaWAN can help to increase the network scalability of LoRa networks by decreasing the number of collisions among signals coming from multiple EDs hence increasing their ability to co-exist.} The network capacity of the LoRaWAN can also be increased by increasing the number of LoRa gateways and reducing the overhearing of transmissions to other gateways by their strategic placement.

\textcolor{black}{In LoRaWAN, a communication channel is observed by the ED and in case of the channel being busy, the PHY-layer parameters (specifically the spreading factor (SF) and the channel frequency) are adjusted reactively. The SFs are partially orthogonal and EDs using different SF values can transmit simultaneously \cite{croce2017impact}.} However, such a reactive approach is not appropriate for low-power EDs because before any parameters selection/adjustment algorithm is invoked, several packets would have been re-transmitted or lost. Moreover, \textcolor{black}{the time delay inherent in the reactive approach} is also not acceptable in situations where decision making has to be done in a bounded time. Hence, there is an absolute need for a proactive, intelligent, and adaptive PHY-layer transmission parameters adjustment algorithm for LoRaWAN.

The presence of a large number of IoT devices increases both the intra-network and inter-network interference causing a performance drop \cite{reynders2016range}. The integration of cognitive radio technology into the LoRaWAN standard can significantly reduce the energy consumption and increase the network capacity \cite{adelantado2017understanding}. A LoRaWAN gateway can decode multiple simultaneous transmissions based on different PHY-layer transmission parameters. Moreover, existing research focuses on a static association between the resources of the IoT and the surrounding real environment. IoT is extremely dynamic in nature and may experience unpredictable mobility, resulting in sudden variations of communication capabilities.

Also, the inherent broadcast nature of wireless communications makes them vulnerable to inter-network interference and adversarial attacks. Jamming of a wireless signal involves the addition of noise to a signal to decrease the signal to noise ratio. It differs from the normal interference in terms of its purpose. LoRaWAN uses encryption techniques that only secure the packet content leaving the transmissions vulnerable. \cite{butun2018analysis}, \cite{sundaram2019survey}, and \cite{reynders2016range} discuss the susceptibility of LoRa networks to jamming attacks. These attacks can lead the resource-constrained IoT devices to: (i) drain their batteries due to repeated data transmissions \cite{namvar2016jamming}; (ii) denial-of-service (DOS).

As the networks are anticipated to become denser in the coming years, both the frequency jamming and dynamicity problems will become more severe. In case of being deployed in real-networks, the performance (PDR and energy consumption) of LoRa network is also affected by interference coming from other deployed networks in the area. This inter-network interference can cause severe performance drop if not managed. There is a need for algorithms that can sense this performance drop and hence adjust the frequencies to minimize the effect of inter-network interference.

Adaptive selection of the PHY-layer parameters in dense LoRa networks can be performed using efficient algorithms hence enabling collision-free concurrent transmissions \cite{latif2017artificial}. The intelligent selection of parameters not only reduces the impact of frequency jamming attacks but also causes a significant drop in energy consumption because of fewer re-transmissions required due to lost or collided packets. For this purpose, we proposed a deep reinforcement learning (DRL)-based PHY-layer parameters selection scheme for dense LoRa networks in our previous work \cite{ilahi2020loradrl}. \textcolor{black}{In that article, we evaluated our proposed technique in uniformly distributed scenarios, with different percentages of intelligent devices, and with different power levels. We showed our technique to be not only able to achieve a high PDR but also reduce energy usage.}

\textcolor{black}{In this paper, we build upon our previous work \cite{ilahi2020loradrl} by increasing its ability to support multiple channel frequencies and perform extensive additional experiments in dense networks containing mobile EDs.} \textcolor{black}{Furthermore, we test our algorithm in case of large-scale jamming attacks on dense networks and show our algorithm to adapt against such attacks.}


The rest of this paper is organized as follows. In Section \ref{sec:background}, we have discussed the common terminologies used in LoRa networks and have provided the related work. In Section \ref{sec:setup}, we have provided the complete system setup of the network and the DRL algorithm. In Section \ref{sec:LoRaDRL}, we have provided a brief introduction of our previously proposed scheme LoRaDRL along with discussing the computational complexity and the applicability to real environments. In Section \ref{sec:experiments}, we have performed the performance evaluation of LoRaDRL in multiple scenarios and provided the multi-channel scheme. Finally, the paper is concluded in Section \ref{sec:conclusion}. A list of important acronyms used are given in Table \ref{tab:acronyms}.


\begin{table}[]
\centering
\caption{Important Acronyms used in the paper}
\label{tab:acronyms}
\begin{tabular}{|c|c|}
\hline
BW & Bandwidth \\ \hline
CR & Code Rate \\ \hline
CSMA/CA & Carrier-Sense Multiple Access / Collision Avoidance \\ \hline
DDQN & Double Deep Q-learning Network \\ \hline
DL & Deep Learning \\ \hline
DNN & Deep Neural Network \\ \hline
DQN & Deep Q-Network \\ \hline
DRL & Deep Reinforcement Learning \\ \hline
ED & End-Device \\ \hline
IoT & Internet of Things \\ \hline
ISM & Industrial, Scientific, \& Medical \\ \hline
LoRa & Long-Range \\ \hline
LoRaWAN & Long-Range Wide Area Network \\ \hline
MAB & Multi-Armed Bandits \\ \hline
MAC & Medium Access Control \\ \hline
PDR & Package Delivery Ratio \\ \hline
PHY & Physical \\ \hline
ML & Machine Learning \\ \hline
RL & Reinforcement Learning \\ \hline
SF & Spreading Factor \\ \hline
WAN & Wide Area Network \\ \hline
\end{tabular}%
\end{table}

\begin{figure*}[!ht]
  \centering
  \includegraphics[width=0.85\linewidth]{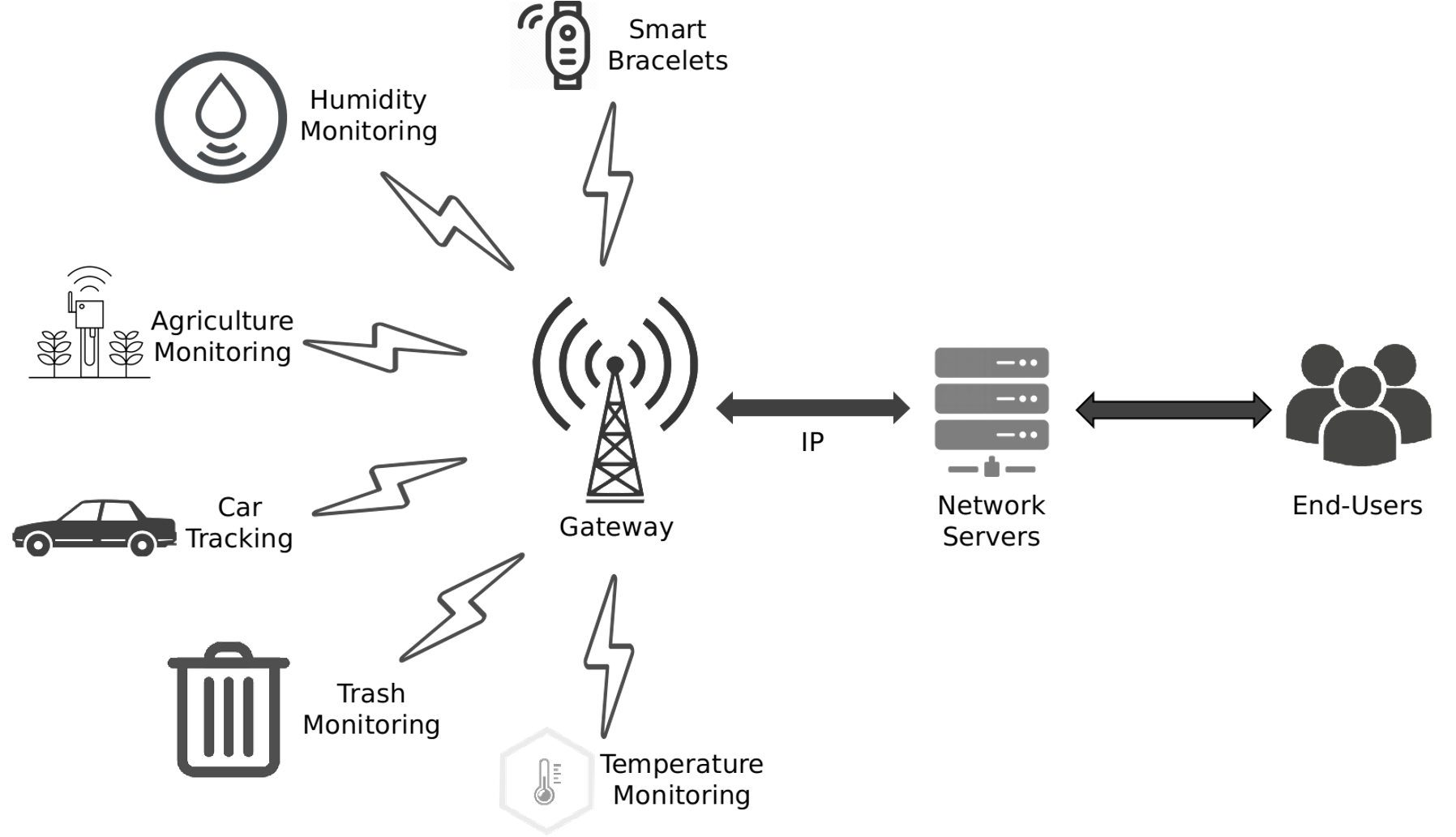}
  \caption{Architecture of LoRaWAN consisting of LoRa EDs, LoRa gateway, network server, and end-user. The EDs serve different purposes and transmit the data to the gateway based on the application requirements. The received data packets are forwarded to the network server, which in turn forwards them to the end-user.}
  \label{fig:lora_architecture}
\end{figure*}

\section{Background and Literature Review}
\label{sec:background}

\subsection{LoRa Networks}

LoRaWAN is laid out in a star-of-stars topology. It works in the unlicensed Industrial, Scientific, and Medical (ISM) frequency band. LoRaWAN architecture consists of LoRa end devices (EDs), LoRa gateways, network servers, and application (user) servers. A basic LoRaWAN architecture containing transmitting EDs, gateway, and network server has been shown in Fig. \ref{fig:lora_architecture}.

\textcolor{black}{The applications of deployed LoRa EDs in LoRaWAN can be either event-driven or scheduled.} The former one involves the transmission of data whenever a specific event occurs while the latter involves transmissions on scheduled intervals. Parking sensors in parking lots to sense the available parking spaces are an example of event-driven EDs while the temperature sensors mounted at the top of buildings to measure the temperature of the area exemplify scheduled transmitters.

In LoRa, a transceiver can select a bandwidth (BW) in the range 7.8 to 500 kHz, and mostly a LoRa transceiver operates at 125 kHz, 250 kHz, or 500 kHz. Spreading Factor (SF) defines the ratio between the symbol rate and the bandwidth. LoRa provides seven SF rates to choose from (SF6 to SF12). By modifying the SF parameter, we make a tradeoff between the communication range and the data rate. \textcolor{black}{As discussed before that the SF values are partially orthogonal, they can be used to make multiple simultaneous communications possible with minimum collisions.} Coding rate (CR) defines the level of protection against interference. LoRa defines four coding rates: $\frac{4}{5}, \, \frac{4}{6}, \, \frac{4}{7}, \, \frac{4}{8}$. A LoRa radio can transmit between -4 to 20 dBm in 1 dB steps. However, due to hardware limitations, the mentioned range is mostly limited between 2 to 20 dBm
The useful bit rate ($R_{b}$) is given as:
\begin{equation}
    R_{b} = SF \times (\frac{BW}{2^{SF}}) \times CR.
\end{equation}

This shows that the useful bit rate is directly proportional to BW \& CR and inversely to SF. LoRaWAN provides 3 transmission classes to satisfy the requirements of different applications, namely, class A, class B, and class C. Class A is the most energy-efficient class and is normally used in battery-powered devices. In class A, there are two downlink communication slots after each uplink transmission. Class B involves scheduled downlink communication slots and is less energy-efficient than class A. In class C, the downlink communication is always active and hence this class is the least energy efficient. Normally, class C devices are connected directly to the main power. To get a broader view of LoRaWAN technology, we refer the reader to \cite{erturk2019survey, de2017lorawan, raza2017low}.


\subsection{Deep Reinforcement Learning (DRL)}

The rapid evolution of deep learning (DL) and computational technologies have enabled the conventional RL to solve the complex sequential decision problem which is previously deemed impossible due to dimensionality issues. A combination of DL with legacy RL is known as DRL. Mnih et al. \cite{mnih2015human} proposed Deep Q-Networks (DQN), a combination of Deep Neural Network (DNN) and Q-learning \cite{watkins1992q}, as a solution to the computational complexity problem faced by Q-learning in complex environments. Mnih et al. \cite{mnih2015human} also introduced the concept of experience replay and target network to improve the DQN's performance. The Q-values are updated as given in Equation \ref{eq1}.

Equation \ref{eq1} provides a detailed description of DQN. 
\begin{equation}
    Q(s, a) = R(s, a) + \gamma \max{_{a^{'}}} (Q(s^{'}, a^{'})), \label{eq1} 
\end{equation}

where $s'$ is the next state, $a'$ is the next action, $R$ is the reward of a state-action pair, $Q$ is the Q-value of the state-action pair, and $\gamma$ is the discount factor. The policy $\pi$ in DQN is to take the action with the maximum Q-value at a specific state, i.e., $Q^{*}(s,a) =  \max_{a}Q^{\pi}(s,a)$ and is represented by the DNN.


\textbf{\textit{Why DRL?}} In case of normal Q-learning, a Q-table is built to store the Q-values corresponding to each state-action pair. This table can only be built when the state-space and action-space are both discrete. In case any of them is continuous, the size of the table increases exponentially with each possible value of actions and states. DQN \cite{mnih2015human} can support continuous state and action spaces while keeping a fixed size of the model. They approximate the relationship of the state-action pairs and Q-values by the use of deep neural networks (DNNs), thereby removing the requirement of populating tables. DQNs were shown to be over-estimating the Q-values by \cite{van2016deep}. As a remedy, they propose the value estimation be done by the target network instead of the online network. This not only reduces the over-estimation of Q-values but also increases the stability of learning.

\subsection{Related Work}

A number of different PHY-layer parameters selection algorithms for LoRaWAN have been proposed so far. We discuss the state-of-the-art algorithms only. There are two major schemes for handling dense LoRa networks: (i) by scheduling the transmissions, and (ii) by an efficient selection of PHY-layer parameters. \textcolor{black}{In 2016, Kim et al. \cite{kim2017adaptive} argued that the adaptive data rate control used by the LoRaWAN protocol is inefficient as it doesn't see the congestion in the LoRa networks. To improve this shortcoming Kim et al. \cite{kim2017adaptive} used a linear regression model and showed better performance by reflecting congestion in the adaptive data rate control.}

Bor et al. \cite{bor2016lora} proposed a LoRaSim simulator for experimenting with dense LoRa networks using different PHY-layer parameter settings. They use fixed subsets of the PHY-layer parameter combinations to ensure collision avoidance. The only problem with their technique is that it suffers from the problems associated with a rule-based mechanism, i.e., their technique is based on a fixed system model and is not able to adapt to the environment changes which are normal in real networks.

\textcolor{black}{Slabicki et al. \cite{slabicki2018adaptive} proposed an end-to-end network simulator called Flora for LoRa networks and also proposed and validated an adaptive data rate scheme for dynamic selection of link parameters for scaleable and efficient network operation. They showed their technique to increase the network delivery ratio under stable channel conditions while keeping the energy consumption low. They showed that the network delivery rate can be further improved using a network-aware approach, wherein the link parameters are configured based on the global knowledge of the network. They did not consider the collisions among packets which can significantly reduce the packet delivery ratio.}

\textcolor{black}{Bianchi et al. \cite{bianchi2019sequential} presented a ``sequential water-filling'' strategy for assigning spreading factors (SFs) to all LoRa nodes. Their design focused on 1) equalizing the time-on-air in the different SF groups; 2) balancing the spreading factor across multiple gateways; 3) keeping into account the channel capture in LoRa. Their work showed an improvement of 38\% capacity over the adaptive data rate provided by LoRaWAN.}

Abdelfadeel et al. \cite{abdelfadeel2019free} propose FREE, a fine-grained scheduling scheme for reliable and energy-efficient data collection in LoRaWAN. They propose that instead of transmitting the data as soon as it is generated, it is scheduled for fixed time slots which are decided by their algorithm. Although this eliminates the problem of collisions in LoRaWAN, this scheduling solution is not scalable for dense networks as each ED will have to wait for its allocated time slot. On the other hand, our algorithm helps efficiently transmit the data as soon as it is generated at the ED with minimized collisions. This also removes the delay caused by the scheduling scheme proposed by \cite{abdelfadeel2019free} which might be destructive in EDs deployed for time-critical applications.

The LoRa network community has also utilized DRL schemes for automating different tasks, such as load balancing \cite{gomez2018machine} and resource management \cite{hussain2019machine}. Aihara et al. \cite{aihara2019q} proposed a Q-learning aided resource allocation and environment recognition scheme for LoRaWAN with CSMA/CA. They train different deep neural networks (DNNs) for each LoRa ED which is resource-intensive. Our technique is only based on training a single DNN for the whole network. Also, the learning of each of the DNN proposed by \cite{aihara2019q} is selfish and every DNN only focuses on its own reward while our technique focuses on a joint reward of the system. Also, such schemes cannot be deployed in dense networks owing to the computational requirements. Techniques like \cite{aihara2019q} fail when they are tested against adversarial jamming attacks because of the inability to adapt to the changes in the environment.

\textcolor{black}{Farhad et al. \cite{farhad2020mobility} propose a pro-active mobility-aware resource assignment algorithm for LoRaWAN. They propose to update the values of SF and transmission power value on each uplink communication. Their algorithm is not based on learning and hence is bound to fail in real environments where the conditions are different from simulation.}

\textcolor{black}{Aggarwal and Nassipuri \cite{aggarwal2021improving} propose to allocate different SF values to the EDs present in a small range of the gateway. This allocation leads to a better performance of the network by increasing the overall PDR. Their algorithm requires to explicitly provide an SF-allocation ratio for the EDs. As our algorithm (discussed later) is based on reinforcement learning, it does this automatically and there is no need to explicitly provide an explicit SF-allocation ratio. An approach similar to \cite{aggarwal2021improving} has also been proposed by \cite{farhad2020resource}.}

\textcolor{black}{Chinchilla et al. \cite{chinchilla2021collision} propose an algorithm for reducing the collisions in LoRa networks. Their algorithm works by dividing the wireless medium into resource blocks where each research block is based on one SF value and one channel frequency. The objective of their algorithm was only to increase the capacity of the network and they did not consider the energy usage in their algorithm. Furthermore, they do not consider the inter-SF collisions while our algorithm (discussed later) takes all of these things into consideration.}

Ta et al. \cite{ta2019lora} proposed the use of RL for dynamic PHY-layer transmission parameters selection for LoRa-based EDs. They pointed out multiple issues with LoRaSim, for example, using perfectly orthogonal spreading factors. Based on their identified weakness in LoRaSim, they proposed another discrete event simulator named LoRa-MAB. They used the Multi-Armed Bandits technique to solve the collision issue. We identify multiple issues with LoRa-MAB and hence propose our centralized DRL-based algorithm as a solution to these issues. The identified issues with LoRa-MAB are listed below:

\begin{enumerate}
    \item LoRa-MAB is exponentially complex in terms of its computational complexity and hence not feasible for dense LoRa networks. The convergence time of the algorithm is high and is bound to increase with an increase in the count of EDs.
    \item It does not account for the mobility of EDs. This makes it inapplicable in a network consisting of mobile EDs, such as health-care, smart vehicles, aging society, and post-emergency networks.
    \item The focus on optimizing power consumption is not done properly. Due to a missing specialized objective function, EDs have the option of choosing any of the available power levels without particularly focusing on saving power. This random choice does not always lead to the optimal power level selection.
    \item The computations are being performed at the EDs without considering the power limitations in the case of battery-powered EDs.
    \item To reduce the complexity of the problem, LoRa-MAB reduces the action space of individual EDs based on their distance from the gateway. In case the EDs are mobile, a change in their position makes the learning sub-optimal.
\end{enumerate}

\textcolor{black}{We refer the reader to \cite{kufakunesu2020survey} and \cite{lehong2020survey} for getting a comprehensive review of the several adaptive resource allocation schemes proposed for LoRaWAN.}

In our previous work \cite{ilahi2020loradrl}, we showed that the performance deteriorates in a LoRa-MAB based system when EDs are mobile.
In this paper, we perform further experiments with LoRaDRL \cite{ilahi2020loradrl} and show the applicability of LoRaDRL to real LoRa networks. Furthermore, we test the performance of LoRaDRL in case of large-scale jamming attacks and show its adaptability to changes in the environment. We also show the susceptibility of rule-based techniques against these attacks.

\section{System Setup}
\label{sec:setup}

We have previously described the working of our DDQN-based adaptive PHY-layer parameter selection algorithm for dynamically deployed networks in \cite{ilahi2020loradrl}. In this paper, we further discuss the complexity and applicability aspects in detail in the following section. One of the major problems seen in the previously proposed resource allocation techniques for LoRaWAN is the missing support for real dynamic environments which keep on changing with time. Furthermore, in the experiments section we show that LoRaDRL can sense the performance drop due to frequency jamming and hence can shift the system to the less interfered channels and hence maintain the performance of the network. We also show the ineffectiveness of the rule-based system LoRaSim against such attacks. 

\subsection{Problem Statement}
LoRa provides multiple SF values for transmission which lead to different data rates. \textcolor{black}{The signals generated using different SF values are partially orthogonal to each other.} By compromising the data rate, the concurrent transmissions can be increased by the use of different SF values and transmission channels. The efficient selection of these parameters in dense networks can not only save energy but also increase the capacity of the network. 

\subsection{System Model}

In our proposed scheme, we consider a single-gateway LoRa network containing $k$ LoRa EDs uniformly distributed over an area of a radius of 4500m with the gateway present in the center. The EDs can choose to transmit the data using different PHY-layer parameter combinations over multiple available transmission channels. \textcolor{black}{We do not limit the SF values for the EDs and all the EDs are free to use any of the SF values.} The gateway acts as the agent whose goal is to decide the PHY-layer parameters for each of the EDs. It is assumed that a new ED arrives at each time-step and is located at an arbitrary location. The normalized count of each of the actions (taken until the current step) and the approximate distance of the new-coming LoRa ED is taken as the state of the environment.

\begin{figure}[!ht]
  \centering
  \includegraphics[width=0.93\linewidth]{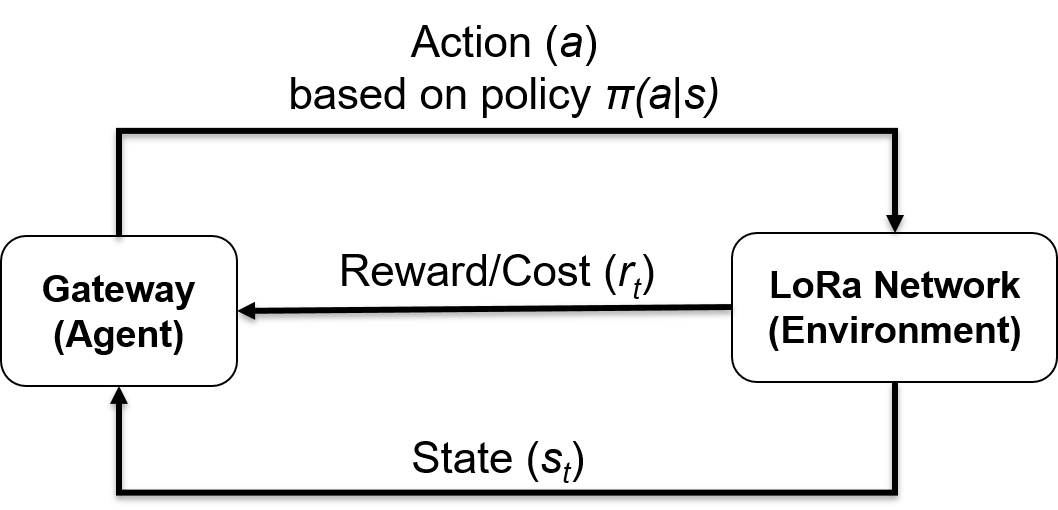}
  \caption{Mapping of our problem setup on DRL. The gateway is considered the agent and the LoRa network represents the environment.}
  \label{fig:loRa_on_DRL}
\end{figure}

The basic mapping of our algorithm on the workflow of DRL has been given in Fig. \ref{fig:loRa_on_DRL}. The agent takes a specific action (choosing a PHY-layer parameters combination for the new LoRa ED) at a time-step and receives a reward based on the achieved packet delivery ratio (PDR) and power-usage based on that chosen action.

We use PDR and energy consumption as our performance metrics. The PDR is defined as the ratio of correctly delivered messages to transmitted messages over a period of time. The achievable PDR depends on the position, count, and behavior of LoRa EDs.

\textcolor{black}{For experimental purposes, we are assuming a LoRa network consisting of class C devices for training. After training our model can be deployed with any class of the LoRa devices.} We use the Gauss-Markov Model for the mobility of the EDs. This model eliminates the sudden stops and sharp turns encountered in the Random Walk Mobility Model by allowing past velocities and directions to influence future velocities and directions.

Our previous results \cite{ilahi2020loradrl} showed that when the mobility was introduced in the LoRa-MAB system \cite{ta2019lora}, the performance started to degrade immediately due to non-adaptability and no support for mobile EDs. The learning in LoRaDRL is performed on the gateway which is independent of the EDs and hence can handle the mobility of EDs. As far as state calculation is concerned, our states are based on the actions taken by the agent until the current step and the approximate distance of the ED from the gateway. The former one can be easily calculated by populating a table while the latter can be approximated using the received power of the signals from the EDs. \textcolor{black}{It is to be noted that the EDs present near buildings will show less received power than in case of open space. This will lead the algorithm to choose higher SF values for such devices which is a good choice.}

The LoRa devices can support a power level as low as 2 dB. In case multiple power level choices are included in the action-space, the energy consumption can reduce considerably. \textcolor{black}{Due to the high training/convergence time of LoRaDRL, we propose the training of LoRaDRL to be performed in simulation and then the model be deployed in real networks with the learning to be continued with a small learning rate. This will help our proposed model adapt and fit itself to the real environment.}

\textcolor{black}{It is to be noted that we assume that the number of EDs and packet arrival rate is known at the gateway. It takes around 5 packet transmissions for the gateway to get a reliable estimate of PDR. For mobile EDs, we assume that the current PDR is being averaged with the previous 4 transmissions’ PDR. As we are not assuming very high velocities, hence the time required to get a good estimate of PDR is acceptable.}

\section{Proposed Scheme for Adaptive PHY-Layer Parameter Selection}
\label{sec:LoRaDRL}

\subsection{Reward Function}
To assist in the learning process, we have designed a specialized reward/cost function to optimize PHY-layer transmission parameters selection for LoRa EDs. By using this reward function, the maximum reward is given to the optimal combination of PHY-layer parameters. The reward function is given in the below equations. Equation \ref{eq3} is the reward for optimizing the PDR of the network only. Equation \ref{eq4} is the modified equivalent to include power optimization.


\begin{equation}
  r_{t} = \alpha*PDR_{ED}- \beta*airtime_{ED}
\label{eq3}
\end{equation}
\begin{equation}
  r_{t} = \alpha*PDR_{ED} - \beta*airtime_{ED} + \gamma*Power_{ED}
\label{eq4}
\end{equation}

where,
\begin{equation}
  Power_{ED} = \frac{Power_{Max} - Power_{Chosen}}{Power_{Max} - Power_{Min}}.
\end{equation}

$ED$ is the new-coming ED that has arrived in the previous time-step, $PDR$ is the package delivery ratio, and $airtime_{ED}$ is the airtime of the specific ED in seconds. $\alpha$, $\beta$ \& $\gamma$ are the relative constants used to assign appropriate weights to $PDR$, $airtime_{ED}$, and $Power_{ED}$. These constants act as hyper-parameters and can be chosen depending on the dynamics of the LoRa network. $Power$ is the reward based on the power choice for the ED. This part of the reward function is designed in such a way that if we have 3 available power levels 3 dB, 6 dB, and 12 dB, the reward is also defined in a distributed fashion. In this way, more reward say $4x$ will be given to the agent if it chooses 3 dB power, lesser reward $3x$ will be given if it chooses the 6 dB power, and the least reward $x$ will be given if it chooses the 12 dB power.

\subsection{Proposed Algorithm}

The proposed DDQN-based algorithm for learning the PHY-layer transmission parameters for EDs in a LoRa network has been given in Algorithm \ref{algo:algo1}. Q-network structure is taken as input to the model and it returns a trained DDQN network at the output. \textcolor{black}{The algorithm trains for a given number of episodes where each episode is run for time-steps equal to the maximum number of EDs present in the system (we assume that a new ED arrives at each time-step). The replay buffer is populated by the agent by taking different actions at different states. Samples from this replay buffer are then used to train the neural network.} This trained network provides the optimal policy for determining the best PHY-layer parameters for the EDs based on the state of the environment.

\begin{algorithm}[!ht]
\caption{LoRaDRL}
\label{algo:algo1}
\textbf{Input:} Q-Network Structure\\
\textbf{Output:} Trained Q-Network\\
\begin{algorithmic}[1]
\STATE Initialize both the Target \& Online Q-Networks
\STATE Initialize the memory (replay buffer)
\FOR{$maxEpisodes$}
    \WHILE{$steps<maxEdCount$}
        \STATE Initialize the LoRa Network
        \STATE Compute state of the Network $s_t$
        \STATE Feed the state to the DNN to get action $a_t$
        \STATE Taken action $a_t$ at state $s_t$
        \STATE Simulate the environment
        \STATE Compute reward $r_t$ and next state ${s}_{t+1}$ 
        \STATE Collect $m$ data-points ${(s_t,a_t,{s}_{t+1},r_t)}$ using policy $\pi$ and add it to the memory
        \STATE Sample mini-batch from memory
        \STATE Compute the change in values using target Q-network $Q_\phi'$: $y_j = r_j + \gamma \max_{a_{j}'} Q_{\phi'}(s_j',a_j')$  
        \STATE Update the Online Q-Network: $\phi \gets \phi-\alpha\sum_j$ $\frac{dQ_\phi(s_j,a_j)}{d\phi}(Q_\phi(s_j,a_j)-y_j)$
        \IF {$steps > targetUpdateInterval$}
        \STATE Update the Target Q-Network $\phi^{'}$
        \ENDIF
        \STATE $s_t \gets s_{t+1}$
    \ENDWHILE
\ENDFOR
\end{algorithmic}
\end{algorithm}

\begin{figure}[!ht]
  \centering
  \includegraphics[width=0.90\linewidth]{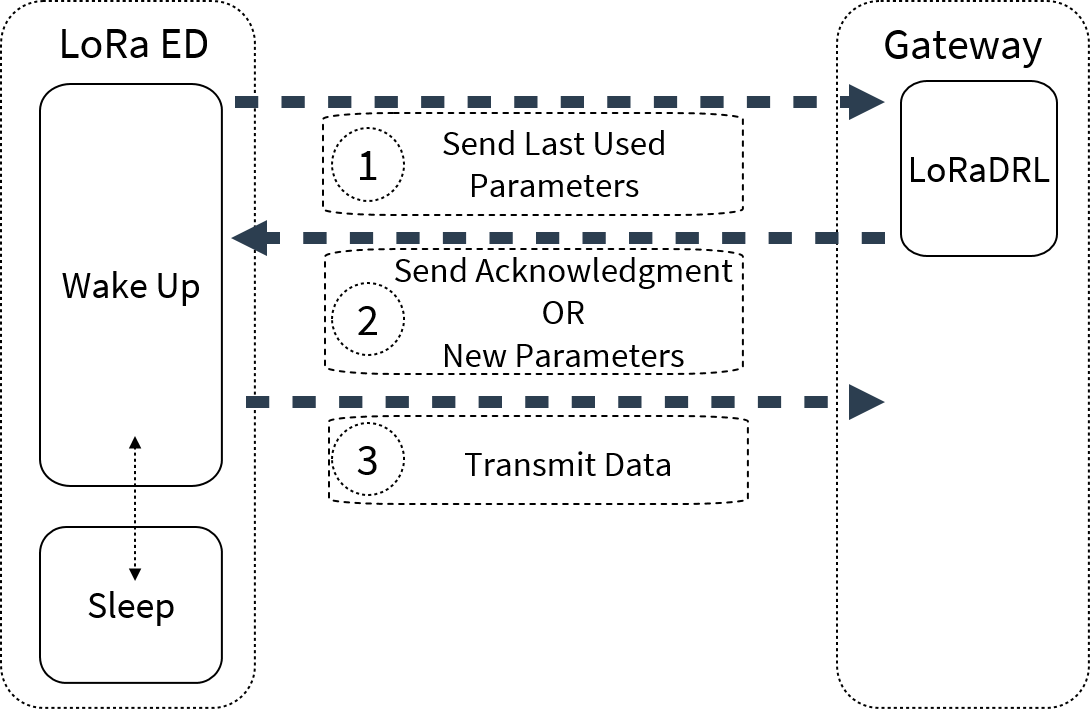}
  \caption{Proposed mechanism for implementing our proposed algorithm in real LoRa networks.}
  \label{fig:LoRa_DRL}
\end{figure}

LoRa EDs are sleeping except when they need to transfer the data. The transmissions are carried out on the base of the different transmission classes, i.e., class A, B, \& C. We propose that the LoRa ED send the packets to the gateway, which in return either sends an acknowledgment (to use the previous parameters) or sends the new PHY-layer parameters combination, to be used for carrying out further transmissions, through the control packets \textcolor{black}{using the fixed bandwidth channel of 125 kHz}. In case the LoRa ED does not receive the parameters or acknowledgment from the gateway, \textcolor{black}{it either chooses the maximum available power and SF to transmit the signal or uses the last allocated PHY-layer parameters for the transmission.} This mechanism has also been shown in Fig. \ref{fig:LoRa_DRL}. Our algorithm also works well on the reduced action space by allowing the agent to choose from a specific subset of actions. This reduced subset can be made according to the data-rate requirements of different applications by fixing a certain SF, CR, or transmission channel. 

\subsection{DRL Specifications}

The neural network is kept small to make the solution more practical. A discount factor of 0.7 has been used to ensure the dependence of the current action on future rewards. Furthermore, we also use the $\epsilon$-greedy learning procedure to fully explore the state-space where action at time-step $t$ is given as:
\begin{align}
a_{t} = \left\{ \begin{array}{cc} 
                {\text{max}}_{a} Q_{t}(s,a) & \hspace{4mm} \text{with probability}\:(1-\epsilon) \\
                \text{Random Action} & \hspace{4mm} \text{with probability}\:\epsilon \\
                \end{array} \right.
\end{align}

We have chosen linear activation at the output layer so that we get a probability for each of the actions. In this way, if some of the EDs can choose only a subset of actions, then that ED can choose the action with the maximum probability from that subset of actions. The target Q-network is updated on regular intervals with the weights of the online network.

\subsection{Computational Complexity}
We have taken the same state-space as our previous submission \cite{ilahi2020loradrl}, i.e., the normalized count of each action and the approximate distance of the new-coming ED from the gateway. Due to this specialized state-space, the complexity of the problem does not increase with the increase of the end-devices. Also, we have chosen a minimal size for the DNN which requires minimal resources for training. In the later sections, we have enhanced our previously proposed scheme to support dense networks. Although the introduction of multiple channels increases both the action and the state count, the same DNN can learn as the goal of the agent in both these cases is the same. The overall complexity of our algorithm is $\mathcal{O}_{NN} + \mathcal{O}(1)$, where \textcolor{black}{$\mathcal{O}_{NN}$ is the complexity of the neural network which is a constant in our case.}

\subsection{Applicability to Real Environments}

We have kept the size of the neural networks the smallest possible. This makes it applicable to gateways being backed by low-end computers. The activation on the last layer has been set to linear. Due to this, our neural network does not train itself to focus on just one action but gives probability to each action in the action-space. So, in case a specific ED is only able to support a sub-space of actions, the ED can choose the possible action with the highest probability.

The DDQN can see the change in the performance of the network based on the reward achieved by taking certain actions in certain states. This ability makes the algorithm adaptive as whenever the DDQN observes a sub-optimal action being performed, it adapts the policy in favor of the better available action. This adaptive behavior is a core benefit of RL. Our proposed centralized approach offers many significant benefits including the ability to adapt (a feature missing in previous solutions) and support for ED's mobility.


\subsection{Multi-Channel Extension of LoRaDRL}
In our previous work, our focus was only on a single-channel and single-gateway scheme. In this work, we have performed a performance evaluation of the scheme under new scenarios and extended the scheme to multi-channel LoRa networks. Modification of the action space is involved in order to include multiple channels to support dense LoRa deployments. We have tested the multi-channel scheme in dense LoRa deployments and shown its ability to manage. We have also tested the performance of LoRaDRL against frequency blocking and shown its ability to adapt to the environmental interference.


\section{Experiments \& Results}
\label{sec:experiments}

For analysis and comparison of our algorithm with the existing state-of-the-art techniques \cite{bor2016lora, ta2019lora}, we perform experiments to evaluate performance under (i) different mobility velocities; (ii) multi-channel dense scenarios; (iii) multiple MAC protocols; and (iv) large-scale frequency jamming attacks. These experiments have been discussed in the following subsections. \textcolor{black}{All of the provided results have just been simulated for experimental purposes. We have made certain design choices just to make it easier for the reviewer to compare the performance of our proposed technique with the other counterparts.}

\textcolor{black}{In our experiments, we use a data frame size of 50 bytes. Typical IoT use cases generate small data packets, hence 50-byte frame size can represent a large number of IoT use cases. In our experiments, the data generation model is based on Poisson distribution as it can model multitude of IoT use-cases’ data traffic generation pattern. We use a mean inter-arrival time, i.e., the average time between two consecutive transmissions of the same ED, of 4 minutes.} \textcolor{black}{The available bandwidth of the LoRa EDs has been fixed to 125 kHz owing to test the performance based on different limitations in different regions. An ED's mean velocity is set to 5 km/h with a variance of 5. We have chosen this velocity to cover the use case of devices mounted on bicycles, UAVs, buildings, etc. for multiple purposes ranging from tracking and transferring sensory data to the central gateway. The specifications of the LoRa simulation have been provided in Table \ref{tab:table2}. The specifications of the neural network have been provided in Table \ref{tab:table1}.}

\begin{table}[t]
\centering
\caption{Specifications of the LoRa Network Simulations}
\label{tab:table2}
\begin{tabular}{|c|c|}
\hline
Average Transmission Interval & $1\times10^{4}$ milliseconds \\ \hline
Mean Rate & 4 minutes \\ \hline
Bandwidth & 125 kHz \\ \hline
Radius & 4500 meters \\ \hline
Transmission Class of EDs & C \\ \hline
Number of Base Stations & 1 \\ \hline
Capture Effect & True \\ \hline
Inter SF Interference & True \\ \hline
Simulation Time of 1 Epoch & 50 $\times$ Mean Rate \\ \hline
Velocity & 5 $\pm$ 5 km/h \\ \hline
\end{tabular}
\end{table}

\begin{table}[t]
\centering
\caption{Specifications of the DDQN in LoRaDRL}
\label{tab:table1}
\begin{tabular}{|c|c|}
\hline
No. of Layers & 2 \\ \hline
No. of Neurons & {[}16, 16{]} \\ \hline
Activations & {[}ReLU, ReLU, Linear{]} \\ \hline
Learning Rate & 0.0005 \\ \hline
Memory Capacity & 30000 \\ \hline
Batch Size & 128 \\ \hline
Gamma for Q-Values & 0.7 \\ \hline
Initial Epsilon & 1 \\ \hline
Final Epsilon & 0.05 \\ \hline
Change in Epsilon & 0.00005 \\ \hline
Update Frequency for Online Network & 3000 \\ \hline
\end{tabular}
\end{table}

\subsection{Performance Under Increasing Mobility}

As discussed in the introduction section, the real networks are a combination of mobile and non-mobile EDs. The mobile EDs move with varying velocities between low and high. In this subsection, we perform experiments to show the ability of LoRaDRL to manage \textcolor{black}{such uniformly distributed heterogeneous networks}. We consider a network of 100 EDs and a single frequency channel available for transmission. The EDs have only a single power level to choose from, i.e., 14 dB. Different velocities of mobile EDs were chosen, i.e., 5 $\pm$ 3 km/hr, and 30 $\pm$ 10 km/hr. The former relates to the health monitoring devices like smartwatches etc communicating with the gateway while the latter relates to EDs mounted on bicycles, carts, etc.  As we are currently considering a network consisting of a single gateway, we have not considered velocities greater than 30 km/hr.

\begin{figure}[!h]
  \centering
  \includegraphics[width=0.93\linewidth]{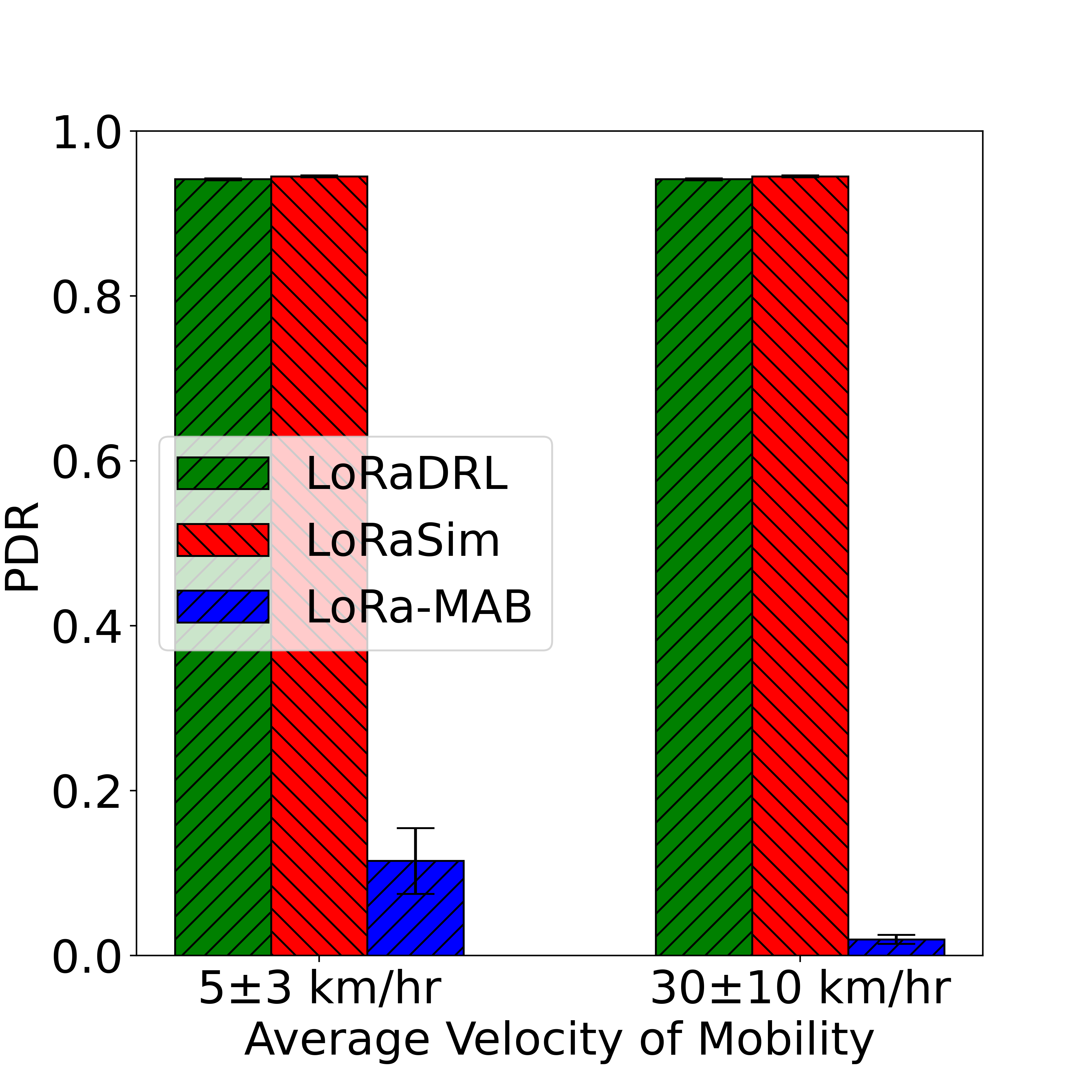}
  \caption{\textbf{Performance evaluation based on different mobility velocities:} Comparison of PDR of LoRa networks under LoRaSim, LoRa-MAB, and LoRaDRL with a confidence interval of 95\%. It can be seen that mobility does not affect the performance of LoRaDRL \& LoRaSim while the increase in velocity causes a deterioration in the performance of LoRa-MAB.}
  \label{fig:mobility_der}
\end{figure}

Fig. \ref{fig:mobility_der} shows the performance of LoRaDRL, the rule-based algorithm proposed by Bor et al. \cite{bor2016lora} and the decentralized algorithm LoRa-MAB. It is visible that the performance of LoRa-MAB drops more with velocities while LoRaSim and LoRaDRL can keep the performance at the same level. \textcolor{black}{The performance of LoRa-MAB drops because of the slow learning process.} However, our proposed PHY-layer parameters selection algorithm can support LoRa networks without any dependence on mobility velocities. 

\subsection{Performance In Multi-channel Scenarios}
\label{sec:channel}

As we know, the spreading factors from SF7-SF12 are partially orthogonal and transmissions with different SFs can be received on the same channel concurrently. Similarly, the frequency channels are also orthogonal and the same SF can be received on different channels without any inter-channel collisions. Current LoRa gateways can receive transmissions from LoRa devices on 8 different channels simultaneously. For this purpose, multi-channel transceivers are used in the LoRa gateway. Different frequencies do not interfere with each other hence the devices can choose from the available SFs without compromising on the PDR.

In our previous work, we had taken the combination of SF and power as the action of the agent. For converting into a multi-channel scheme, we add the channel frequency to the action space hence increasing the action count according to the available frequencies. 
For testing the performance of LoRaDRL in dense deployments, we consider an environment consisting of 1000 LoRa EDs and a single gateway. The available choices of frequency channels are set to 8 which is the maximum number of frequency channels a LoRa gateway can receive and decode simultaneous transmissions. The EDs have only a single power level to choose from, i.e., 14 dB.

The node-count wise PDR of LoRaDRL during learning in dense LoRa networks has been shown in Table \ref{tab:der_dense_stats}. The results show that our model can manage these networks effectively. In the table, a very small drop of PDR can be seen with the increase in the count of EDs. \textcolor{black}{Table \ref{tab:der_dense_SF}(a) shows the percentage of SF values allocated to the devices in this dense network across all the frequencies. Table \ref{tab:der_dense_SF}(b) shows the per-SF PDR performance of the LoRa network.} \textcolor{black}{These tables show that LoRaDRL allocates the PHY-layer parameter values dynamically and adaptively. On the other hand, LoRaSim and LoRa-MAB allocate the values based on the distance from the gateway.}



\begin{table}[!ht]
\caption{Table showing the performance of an 8-channel LoRaDRL in a dense LoRa network consisting of a single base-station. The values are presented with 95\% confidence interval.}
\label{tab:der_dense_stats}
\centering
\begin{tabular}{|c|c|}
\hline
No. of Nodes & DER \\ \hline
250 & $0.94 \pm 0.0091$ \\ \hline
500 & $0.91 \pm 0.0093$ \\ \hline
750 & $0.88 \pm 0.0093$ \\ \hline
1000 & $0.83 \pm 0.01$ \\ \hline
\end{tabular}
\end{table}


\begin{table}[!ht]
\caption{(a) Table showing the percentage of SF values of an 8-channel LoRaDRL in a dense LoRa network consisting of a single base-station. (b) Table showing the per-SF PDR performance of an 8-channel LoRaDRL in a dense LoRa network consisting of a single base-station.}
\label{tab:der_dense_SF}

\vspace*{0.3 cm}

\begin{subtable}[!ht]{\linewidth}
\centering
\begin{tabular}{|c|c|}
\hline
SF-value & \begin{tabular}[c]{@{}c@{}}Percentage of Devices \\ Allocated\end{tabular} \\ \hline
SF-7 & 4.93 \\ \hline
SF-8 & 15.09 \\ \hline
SF-9 & 23.4 \\ \hline
SF-10 & 19.73 \\ \hline
SF-11 & 18.38 \\ \hline
SF-12 & 18.47 \\ \hline
\end{tabular}
\vspace*{0.1 cm}
\caption{}
\end{subtable}

\vspace*{0.3 cm}

\begin{subtable}[!ht]{\linewidth}
\centering
\begin{tabular}{|c|c|}
\hline
SF-value & Per-SF PDR Performance \\ \hline
SF-7 & 0.99 \\ \hline
SF-8 & 0.98 \\ \hline
SF-9 & 0.79 \\ \hline
SF-10 & 0.83 \\ \hline
SF-11 & 0.8 \\ \hline
SF-12 & 0.76 \\ \hline
\end{tabular}
\vspace*{0.1 cm}
\caption{}
\end{subtable}

\end{table}





\subsection{Performance With Different MAC Protocols}

\begin{figure}[!t]
  \centering
  \includegraphics[width=0.93\linewidth]{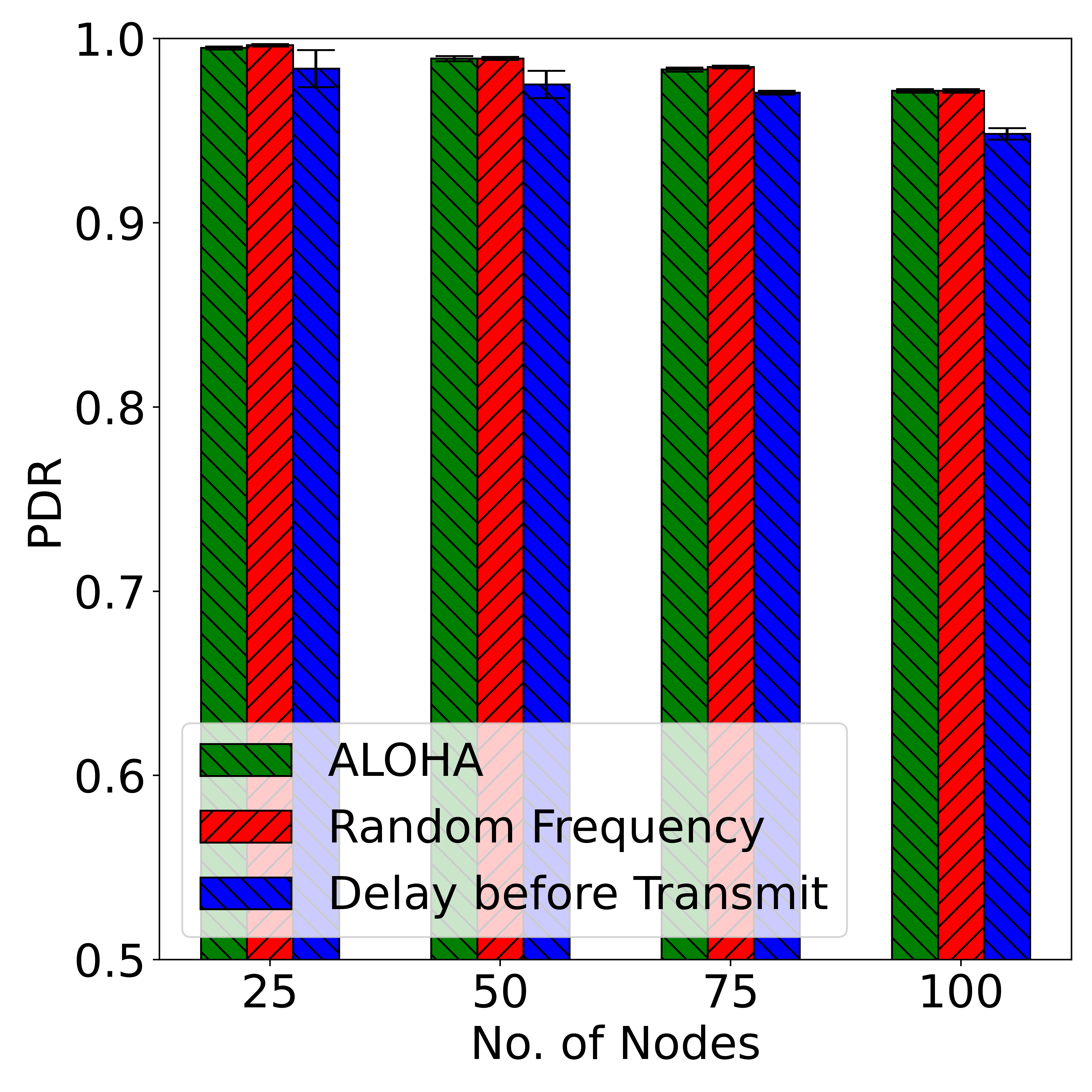}
  \caption{Figure showing the performance of LoRaDRL with different MAC protocols. It can be seen that there is a minor performance difference while using different MAC protocols with LoRaDRL. The bars are plotted with 95\% confidence interval.}
  \label{fig:pdr_mac_protocol}
\end{figure}

Much previous work has been focused on improving the LoRa performance using different MAC protocols than pure ALOHA. By using our proposed algorithm LoRaDRL, we can use the basic ALOHA to perform similar to complex MAC protocols. In this subsection, we test the performance of our proposed algorithm LoRaDRL with multiple MAC protocols. We consider a 2-channel LoRa network containing 100 uniformly placed EDs. The delay in case of ``delay before transmit'' is calculated using the following equation:

\begin{equation}
    T_{D} = (ED_{ID} \times U_{d})\:mod\:Pkt_{iat},
\end{equation}

where $ED_{ID}$ is the ID of the respective ED, $U_{d}$ is the delay in microseconds, and $Pkt_{iat}$ is the node mean packet arrival time. For these experiments, $U_{d}$ was set to 1000. Fig. \ref{fig:pdr_mac_protocol} shows the observed performance. Only a minor difference in performance can be seen as all the features of these MAC protocols are already present in LoRaDRL. Furthermore, LoRaDRL reduces the burden on ordinary nodes by pushing the complexity to a central entity (the gateway). The performance of LoRaSim with different MAC protocols has been shown in \cite{farooq2018search}.

Channel sensing multiple access (CSMA) involves the sensing of the channel before transmission and transmitting if the channel is free else waiting for a certain time interval and then sensing the transmission channel again. In this way, the EDs have to wait for the channel to become free which is a rare case in dense networks. LoRaDRL enables concurrent data transmissions and removes the requirement of sensing the channel and waiting. This reduces the power requirement for the EDs and shifts the complexity from the resource-constrained EDs to the gateway. \textcolor{black}{Furthermore, most of the state-of-the-art MAC layer protocols for LoRa are complex while LoRaDRL is based on ALOHA.}

\subsection{Performance Under Adversarial Frequency Jamming Attacks}

\textcolor{black}{Large-scale frequency jamming attacks can be avoided by a continuous shifting of frequencies hence making the jamming difficult \cite{aras2017selective}. However, in realistic settings, the presence of an intermediary to continuously change the settings is not necessary. This puts forward the need for intelligent algorithms that can adapt to the changing environment in the favor of optimal settings. Our technique can adapt to jamming attacks and can retain the performance of the LoRa network by frequency hopping. In the case of RL, the learning and prediction go hand in hand which makes it proactive to adversarial attacks and adaptive to the changing conditions \cite{restuccia2018securing}.}

For this experiment, we assume that there is another network present in the area who is generating very high inter-network interference hence reducing the performance of our (LoRa) network. We consider a network consisting of 100 LoRa EDs and two frequency channels available for transmission. The EDs have only a single power level to choose from, i.e., 14 dB. The network is taken to be uniformly distributed with the EDs moving with random velocities under 1 km/hr.

\begin{figure}[!t]
    \centering
    \includegraphics[width=0.93\linewidth]{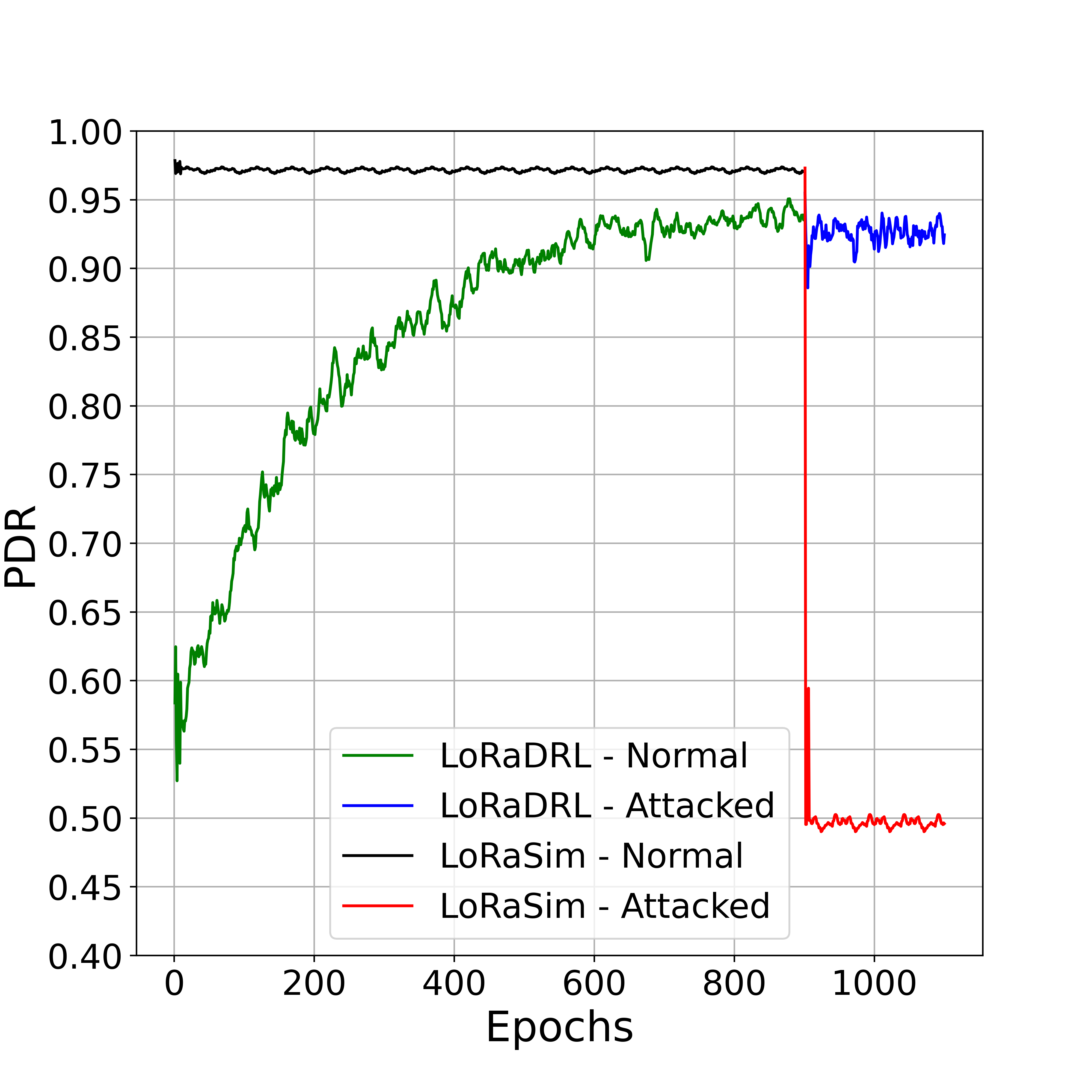}
    \caption{Figure showing the performance of a multi-channel LoRaDRL and LoRaSim scheme under frequency jamming attack. A small drop of performance is seen in the case of LoRaDRL as it shifts to the other available frequency. While in the case of LoRaSim, a sudden drop of performance can be seen because of the absence of a feedback loop.}
    \label{fig:der_2channel_freqblock}
\end{figure}

Fig. \ref{fig:der_2channel_freqblock} shows the training of multi-channel LoRaDRL algorithm. At epoch 900, one frequency out of the two available ones is jammed. This jamming results in a sudden drop in performance. The system later learns on the base of the current performance and can adapt to the changing environment and achieve the performance of single-channel LoRaDRL. While in the case of a frequency jamming attack on a rule-based LoRaSim, the performance drops to half. The reason for this is the random selection of a frequency channel for each transmission. The performance of LoRaSim under frequency jamming attack has been shown in this figure which shows no retention of performance. \textcolor{black}{It is to be noted here that the collisions and jamming with respect to the downlink communication is left as future work.}

\vspace{3mm}

\textcolor{black}{All of the provided experiments were performed on a low-end 4th generation i3 laptop. It took on average 0.3s for LoRaDRL to make a decision while LoRaSim took on average 0.01s to make a decision. An important aspect is the learning ability of the LoraDRL based on the changes in the environment whereas loraSim assigns parameters based on a defined set of rules.}

\section{Conclusions}
\label{sec:conclusion}
\textcolor{black}{We have proposed an intelligent multi-channel resource allocation algorithm for dense LoRa networks termed as LoRaDRL.} We have provided a detailed performance evaluation of this proposed algorithm by testing it in LoRa networks consisting of LoRa end-devices (EDs) having different mobility velocities, and in dense LoRa deployments. Our scheme has shown exceptional results when compared with similar previous techniques. Furthermore, we have proposed a multi-channel scheme for LoRaDRL to support multiple channels. We tested the performance of LoRaDRL with different MAC protocols and show its ability to manage the system while shifting the complexity from the EDs to the gateway. We have also tested our proposed technique under large-scale jamming attacks where the rule-based techniques fail badly. The results show the effectiveness of our proposed technique against such attacks and its adaptiveness to the changes in the environment.

\bibliographystyle{IEEEtran}
\balance
\bibliography{references}

\end{document}